# 動的なデバイス連携サービスの連携ロジック記述と実行の検討

A study of coordination logic description and execution for dynamic device coordination services

山登庸次[1]　干川尚人[1]　野口博史[1]　出水達也[1]　片岡操[1]
Yoji Yamato　Naoto Hoshikawa　Hirofumi Noguchi　Tatsuya Demizu　Misao Kataoka

日本電信電話株式会社　NTT ネットワークサービスシステム研究所[1]
NTT Network Service Systems Laboratories, NTT Corporation

## 1 はじめに

近年 IoT 技術 [1]-[11] やクラウド技術 [12]-[33] が進展し，数多くのデバイスがネットワークに接続され活用できるようになっている．多彩な IoT サービスをコスト低く開発，運用するためには，デバイスとサービスを分離し，水平分離的にデバイスとサービスを相互に利用可能にすること（オープン IoT の考え）が必要となる．

私達は，オープン IoT に向けて，サービスからデバイスを自由に利用するための仕組みとして，Tacit Computing（TC）技術を提案している [34]．Tacit Computing は，ユーザが必要なデータを持つデバイスをオンデマンドに発見し，利用する技術である．しかし，複数のデバイスを動的に連携する際にその連携ロジックについては十分検討されていない．そこで，本稿では，動的なデバイス連携の連携ロジック記述と実行について提案する．

## 2 サービス例と課題

サービス例として，駅での外国人観光客グループへのナビを考える．連携するデバイスは，駅のカメラ，ディスプレイ，放送スピーカである．観光客が行先をスマホやタッチパネルで指定すると，駅のディスプレイに乗換先が表示され，駅のスピーカーで案内が流れる．更に，駅のカメラで観光客グループをモニタして，もし誤った方向に向かった場合に，スピーカーでアラートをする．

ここで，各駅で設置されている，ディスプレイ，カメラ，スピーカーは，機種や機器インタフェース（IF）が異なるため，鉄道会社毎に連携するプログラムを書く必要があるのが現状である．そのため，動的なデバイス連携では，連携ロジックは，個々のデバイスに依存せず記述できることが必要である．しかし，連携ロジックの記述と実行について，デファクト標準な方法は無く，課題となっている．なお，デバイス側は，自身が持つ機能と位置，アクセス手段等のメタデータを公開しているとする．

## 3 連携ロジック記述と実行の 3 方式案

案 1：標準インタフェースを使い連携ロジックを記述

UPnP デバイス等，デバイスが持つサービスのアクション（機能）が標準化されているデバイスがある．そのため，UPnP アクション等，標準 IF を使って連携ロジックを記述することで，実行時は連携ロジックを TC サーバ等が解釈し，TC サーバより標準化された手法（SOAP 等）を用いて，デバイスを連携する．

メリット：標準 IF を使うため具体的デバイスに依存せず記述できる．デメリット：標準化されたデバイスしか連携できない．

案 2：連携ロジックは抽象的に記述し，実行時は TC サーバより具体的 IF で連携

デバイスは多種多様であり標準化されていない物も多い．そこで，連携ロジックは，デバイスの具体的 IF に依存しない抽象的な指示として記述する．ユーザからリクエストあった際に，TC サーバは，連携ロジックを取得するとともに，デバイスのメタデータやその時点のデータを元に利用するデバイス群を選択する．選択したデバイス群を連携するサービス実行の際に，抽象的指示と，デバイスメタデータの具体的 IF 間のマッピングを取り，TC サーバより具体的 IF に基づいてデバイスを連携利用する．メタデータ及び抽象的記述として，Semantic Web Services 技術等が候補としてある．

メリット：具体的デバイスに依存しない抽象的記述で連携ロジックを記述できる，TC サーバでサービス実行が完結する．デメリット：メタデータや抽象的記述の標準化や普及が必要（Semantic Web Services 等 [35]-[72]）である，TC サーバでデバイスへの多彩なアクセス手段を準備する必要がある．

案 3：連携ロジックは抽象的に記述し，実行時は TC サーバよりの抽象的指示を GW にて具体的 IF に変換

案 2 同様，連携ロジックは，デバイスの具体的 IF に依存しない抽象的な指示として記述．ユーザからリクエストあった際に，TC サーバは，連携ロジックを取得するとともに，デバイスのメタデータやその時点のデータを元に利用するデバイス群を選択する．選択したデバイス群を連携するサービス実行の際に，TC サーバは抽象的指示をデバイスを収容する GW に標準的手段（REST 等）で送信し，GW で個々のデバイスに依存した具体的 IF に基づいてデバイスを利用する．

メリット：具体的デバイスに依存しない抽象的記述で連携ロジックを記述できる，TC サーバは GW に抽象的指示を標準的手段で送ればよい．デメリット：メタデータや抽象的記述の標準化や普及が必要である，GW 側で抽象的指示からデバイスの具体的 IF アクセスへの変換手段を持つ必要がある．

## 4 提案

案 1 は多様性がない．案 2，3 は一長一短のため，折衷案として，SOAP, REST 等のメジャーな手段で制御できるデバイスは，案 2 の TC サーバより直接連携し，デバイスの制御がマイナーな手段や無線等の場合は，案 3 の TC サーバからは SOAP，REST 等で抽象的指示を送り，GW で変換して連携する形を提案する．